\documentclass[sigconf]{acmart}
\usepackage[draft]{todonotes}
\usepackage{xspace}
\usepackage{enumitem}
\usepackage{booktabs}
\usepackage{pgfplots}
\usepackage{pgfplotstable}
\pgfplotsset{compat=newest}
\usepackage{filecontents}
\usepackage{multirow}
\usepackage[binary-units=true]{siunitx}
\usepackage{amsmath}

\renewcommand\footnotetextcopyrightpermission[1]{} 
\let\savedCaption=\caption
\renewcommand{\caption}[1]{\savedCaption{#1}\vspace{-1.5em}}
\newcommand{\tcaption}[1]{\caption{#1}\vspace{-0.5em}}

\newcommand\sysname{Contour\xspace}

\DeclareSIUnit{\microsecond}{\SIUnitSymbolMicro\second}

\newcommand*\dash{\ifvmode\quitvmode\else\unskip\kern.16667em\fi---%
\hskip.16667em\relax}

\newcommand{\publication}[1]{\noindent\vspace*{-1em}\raisebox{25pc}[0pt][0pt]{\hspace*{-0pt}\noindent\parbox[t]{6.5in}{\sl{#1}}}}

\newenvironment{sarahlist}{
\begin{description}[itemsep=2pt,leftmargin=0.2cm]
}{\end{description}}

\newcommand\authority{\mathsf{Authority}}
\newcommand\auditor{\mathsf{Auditor}}
\newcommand\monitor{\mathsf{Monitor}}
\newcommand\archivist{\mathsf{Archivist}}

\newcommand\tx{\mathsf{tx}}

\newcommand\head{\mathsf{head}}

\newcommand\roothash{\ensuremath{h}}
\newcommand\ourbinary{\mathsf{bin}}

\newcommand\proveinclusion{\mathtt{prove\_incl}}
\newcommand\checkinclusion{\mathtt{check\_incl}}
\newcommand\commit{\mathtt{commit}}
\newcommand\getcommits{\mathtt{get\_commits}}
\newcommand\getarchstate{\mathtt{get\_arch\_state}}
\newcommand\sync{\mathtt{sync}}

\newcommand\paragraphn[1]{\noindent\textbf{#1}}

\begin{document}

\title{\sysname: A Practical System for Binary Transparency}

\author{Mustafa Al-Bassam}
\affiliation{%
  \institution{University College London}
}
\email{m.albassam@cs.ucl.ac.uk}

\author{Sarah Meiklejohn}
\affiliation{%
  \institution{University College London}
}
\email{s.meiklejohn@ucl.ac.uk}

\begin{abstract}

Transparency is crucial in security-critical applications that
rely on authoritative information, as it provides a robust mechanism for
holding these authorities accountable for their actions.  A number of
solutions have emerged in recent years that provide transparency in the
setting of certificate issuance, and Bitcoin provides an example of how to
enforce transparency in a financial setting.  In this work we shift to a new
setting, the distribution of software package binaries, and present a system
for so-called
``binary transparency.''  Our solution, \sysname, uses proactive methods for
providing transparency, privacy, and availability, even in the face of
persistent man-in-the-middle attacks.  We also demonstrate, via
benchmarks and a test deployment for the Debian software repository, that
\sysname is the only system for binary transparency that satisfies the
efficiency and coordination requirements that would make it possible to
deploy today.

\end{abstract}

\maketitle

\publication{Full version of an extended abstract published in \textit{CBT 2018}.}

\section{Introduction}\label{sec:introduction}

Historically, functional societies have relied to a large degree on trust in
their governing institutions, with participants in various systems (nation
states, the Internet, financial markets, etc.) trusting those in charge to
follow an agreed-upon set of rules and thus provide the system with some level
of integrity.
In recent years, however, increasing numbers of incidents have
demonstrated that integrity cannot be meaningfully achieved solely by placing
trust in a small number of entities.  As a result, people are now demanding
more active participation in the systems with which they interact, and more
accountability for the entities that govern them.  The main method
that has been relatively successful thus far in achieving accountability
is the idea of \emph{transparency}, in which information about the decisions
within the system
are made globally visible, thus enabling any participant to check for
themselves whether or not the decisions comply with what they
perceive to be the rules.

One of the technical settings in which the idea of transparency has been most
thoroughly\dash and successfully\dash deployed is the issuance of X.509
certificates.  This is partially due to the nature of these
certificates (which are themselves intended to be globally visible), and
partially to the many publicized failures of major certificate authorities
(CAs)~\cite{dangoodin2012,johnleyden2011}.  A long line of recent
research~\cite{ietf-ct,arpki,aki,enhanced-ct,coniks,ct-eprint,ikp,alintomescu2016}
has provided and analyzed solutions that bring transparency to the
issuance of both X.509 certificates (``certificate transparency'') and to the
assignment of public keys to end users (``key transparency'').

Despite their differences, many of these systems share a fundamentally similar
architecture~\cite{melissachase2016}: after being signed by CAs,
certificates are stored by \emph{log servers} in a globally visible
append-only log; i.e., in a log in which entries cannot be deleted without
detection.  Clients are told to not accept certificates
unless they have been included in such a log, and to determine this they rely
on \emph{auditors}, who are responsible for checking inclusion of the
specific certificates seen by clients.  Because auditors are often thought of
software running on the client (e.g., a browser extension), they must be able
to operate efficiently.  Finally, in order to expose misbehavior,
\emph{monitors} (inefficiently) inspect the certificates stored in a given log
to see if they satisfy the rules of the system.

To prevent clients from accepting bad certificates, such systems thus rely on
monitors to expose them.  Because auditors are the ones
communicating with the client, however, to achieve this
property an additional line of communication is needed between the auditor and
monitor in the form of a \emph{gossip}
protocol~\cite{linusnordberg2015,laurentchuat2015}.  In such a protocol, the
auditor and monitor periodically exchange information on their current and previous views of the log,
which allows them to detect whether or not their views are \emph{consistent},
and thus whether or not the log server is misbehaving by presenting ``split''
views of the log.  If such attacks are possible, then the accountability of
the system is destroyed, as a log server can present one log containing all
certificates to auditors (thus convincing it that its certificates are in
the log), and one log containing only ``good'' certificates to monitors (thus
convincing them that all participants in the system are obeying the rules).

While gossiping can detect this misbehavior, it is ultimately a
retroactive mechanism\dash i.e., it detects this behavior after an auditor
has already accepted a certificate as valid and it is too late\dash and is thus
most effective in settings where (1) no persistent man-in-the-middle (MitM) attack can
occur, so the line of communication between an auditor and monitors remains
open, and (2) some form of external
punishment is possible, to sufficiently disincentivize misbehavior on the basis
of detection. Specifically for (1), if an auditor has no means of communication
that is not under an adversary's control for the foreseeable future (a
scenario we refer to as a persistent MitM attack), then the
adversary may block all gossip being sent to and from the auditor, and thus
monitors may never see evidence of log servers misbehaving.

Such a persistent MitM attack may be performed by an adversary
who has compromised the cryptographic signing keys of the software
distribution authority.  This would enable them to compromise individual
devices with malicious software updates, and then prevent gossiping between
auditors and monitors by either using the malicious software to disable the
gossiping system, or\dash if they control the network the device is connect
to\dash prevent gossiping at a network level until the device stops
gossiping.  For example, the proposed gossip protocol for CT implements a
fixed-sized pool of items to gossip, with items eventually being
removed from the pool, as it would be wasteful for devices to gossip about the
same information permanently~\cite{linusnordberg2015}.  An adversary would
then have to carry out an attack only until this pool were emptied.

Various systems have been proposed recently that use proactive transparency
mechanisms designed to
operate in settings where these assumptions cannot be made, such as Collective
Signing~\cite{ewasyta2015} (CoSi), but perhaps the most prominent example of
such a system is Bitcoin (and all cryptocurrencies based on the idea of a
\emph{blockchain}).  In Bitcoin, all participants have historically
played the simultaneous role of log servers (in storing all Bitcoin
transactions), auditors, and monitors (in checking that no double-spending
takes place).  The high level of integrity achieved by this comes at
great expense
to the participants, both in terms of storage costs (the Bitcoin blockchain is
currently over
\SI{100}{\giga\byte}\footnote{\url{blockchain.info/charts/blocks-size}})
and computational costs in the form of the expensive proof-of-work mechanism
required to maintain the blockchain, but several recent proposals attempt to
achieve the same level of integrity in a more scalable
way~\cite{alintomescu2016,bitcoin-cosi}. CoSi~\cite{ewasyta2015} achieves this
property by allowing a group of witnesses to collectively sign statements to
indicate that they have been ``seen,'' but assumes the setup and maintenance
of a Sybil-free set of witnesses, which introduces a large coordination
effort.

Because of the effectiveness of these approaches, there has been interest
in repurposing them to provide not only transparency for
certificates or monetary transfers, but for more general classes of
objects (``general transparency''~\cite{continusec}).  One specific area that
thus far has been relatively unexplored is the setting of software
distribution (``binary
transparency'').
Bringing transparency to this setting is increasingly important, as there are
an increasing number of cases in which actors target devices with malicious
software signed by the authoritative keys of update servers.  For example, the
Flame malware, discovered in 2012, was signed by a
rogue Microsoft certificate and masqueraded as a routine
Microsoft software update~\cite{dangoodin2012}.
In 2016, a US court compelled Apple to produce and sign custom firmware
in order to disable security measures on a phone that the FBI wanted to
unlock~\cite{cyrusfarivar2016}.
\\~\\
\paragraphn{Challenges of binary transparency.} Aside from its growing
relevance, binary transparency is particularly in need of exploration because
the techniques described above for both certificate transparency and Bitcoin
cannot be directly translated to this setting.  Whereas certificates and Bitcoin
transactions are small (on the order of kilobytes), software binaries can be
arbitrarily large (often on the order of gigabytes), so cannot be easily stored
and replicated in a log or ledger.

Most importantly, by their very nature software packages have the ability to
execute arbitrary code on a system, so malicious software packages can easily
disable gossiping mechanisms, and we cannot assume that the auditor always
has a means of communication that is not under an adversary's control.
Specifically, as discussed earlier a malicious adversary may perform a
MitM attack to
prevent gossip while presenting an auditor a malicious view of the log, and the
log may itself contain a malicious software update that executes code to disable
gossiping. This makes retroactive methods for detecting misbehavior uniquely
poorly suited to this setting, in which clients need to know that a software
package has been inspected by independent parties \emph{before} installing
it, not after. Binary transparency systems relying on such retroactive methods,
based on Certificate Transparency, are currently being proposed for
Firefox~\cite{mozillabt2017}.
\\~\\
\paragraphn{Our contributions.} We present \sysname, a solution for binary
transparency that utilizes the Bitcoin blockchain to proactively prevent clients
from installing malicious software, even in the face of long-term MitM attacks.
Concretely, we contribute a realistic threat model for this setting and
demonstrate that \sysname is able to meet it; we also show, via comparison with
previous solutions, that \sysname is currently the only solution able to satisfy these
security properties while still maintaining efficiency and a minimal level of
coordination among the various participants in the system.  We also provide a
prototype implementation that further demonstrates the efficiency of \sysname,
and finally provide an argument for its practicality via a test deployment for
the Debian software repository.  Putting everything together, we view \sysname
as a solution for binary transparency that is ready to be deployed today.

We begin in Section~\ref{sec:threat} by presenting our threat model.  In
addition to the goal of preventing split views, we highlight
the importance of \emph{auditor privacy}, in which auditors should not reveal
the particular binaries in which they are interested (as this could reveal,
for example, that a client has a susceptible version of some software), and of
\emph{availability}, in which auditors and monitors should still be able to do
their job even if the original software update server loses its data or goes
offline.

After then presenting the design of \sysname in Section~\ref{sec:design}, we
go on to analyze both its security and its efficiency in
Section~\ref{sec:evaluation}.  Given the volume of related research on
certificate transparency, we also present some comparisons here, and argue
that ours is the first efficient solution to provide these security guarantees
without requiring any coordination cost, in the form of selecting a
central entity to perform authorization, or otherwise trusting some party
to form a Sybil-free set of nodes.  

To validate our efficiency claims, in Section~\ref{sec:impl} we describe an
implementation of \sysname and benchmark its performance, finding that almost
all operations can be performed very quickly (on the order of microseconds),
that auditors can store minimal information (on the order of
kilobytes), and that arbitrary numbers of binaries can be represented by a
single small (235-byte) Bitcoin transaction.  We also
validate our claims of real-world relevance by presenting, in
Section~\ref{sec:debian}, the application of \sysname to the current
package repository for the Debian operating system.  We find that it would
require minimal overhead for existing actors, and
cost under 17 USD per day (even given the current high price of Bitcoin).

Finally, in Section~\ref{sec:discussion} we present some possible extensions
to \sysname, including a discussion of how to use it to achieve general
transparency, and in Section~\ref{sec:conclusions} we conclude.

\section{Related Work}\label{sec:related}

There is by now a significant volume
of related work on the idea of transparency, particularly in the settings of
certificates, keys, and Bitcoin.  We briefly describe some of this work here,
and provide a more thorough comparison to the most relevant work in
Section~\ref{sec:compare-others}.  While \sysname uses similar techniques to
previous solutions within these other contexts, to the best of our knowledge
it is the first full deployable solution in the context of binary transparency.

In terms of certificate transparency, AKI~\cite{aki} and ARPKI~\cite{arpki}
provide a distributed infrastructure for the issuance of certificates, thus
providing a way to prevent rather than just detect misbehavior.  Certificate
Transparency (CT)~\cite{ietf-ct} focuses on the storage of certificates rather
than their issuance, Ryan~\cite{enhanced-ct} demonstrated how to handle
revocation within CT, and Dowling et al.~\cite{ct-eprint} provided a proof of
security for it. Eskandarian et al.~\cite{sabaeskandarian2017} propose how to
make some aspects of gossiping in CT more privacy-friendly using zero-knowledge
proofs. CONIKS~\cite{coniks} focuses instead on key transparency, and thus pays
more attention to privacy and does not require the use of monitors (but rather
has users monitor their own public keys).

In terms of solutions that avoid gossip, Fromknecht et 
al.~\cite{certcoin} propose a decentralized
PKI based on Bitcoin and Namecoin, and IKP~\cite{ikp} provides a way to issue
certificates based on Ethereum.  EthIKS~\cite{ethiks} provides an
Ethereum-based solution for key transparency and Catena~\cite{alintomescu2016}
provides one based on Bitcoin.  While both Catena
and \sysname utilize similar recent features of Bitcoin to achieve efficiency,
they differ in their focus (key vs.\ binary transparency), and thus in the
proposed threat model; e.g., Catena dismisses eclipse
attacks~\cite{atulsingh2006} on the Bitcoin network, whereas we consider them
well within the scope of a MitM attacker. Chainiac~\cite{nikitin2017} is a
system for proactive software update transparency based on a 
verifiable data structure called a skipchain.  Chainiac uses a consensus
mechanism based on Collective Signing (CoSi)~\cite{ewasyta2015}, leading to
the need for an authority to maintain a Sybil-free set of nodes.

Finally, in terms of more general solutions, Chase and Meiklejohn abstract
CT into the general idea of a ``transparency
overlay''~\cite{melissachase2016} and prove its security.  Similarly,
CoSi~\cite{ewasyta2015,bitcoin-cosi} is a general consensus
mechanism that shares our goal of providing transparency even in the face of
MitM attacks and thus avoids gossiping, but requires setting up a distributed
set of ``witnesses'' that is free of Sybils.  This is a deployment overhead
that we avoid.

\section{Background}\label{sec:background}

\subsection{Software distribution}

Software distribution on modern desktop and mobile operating systems is managed
through centralized software repositories such as the Apple App Store, the
Android Play Store, or the Microsoft Store.  Most Linux
distributions such as Debian also have their own software repositories from
which administrators can install and update software packages using command-line
programs.

To reduce the trust required in these repositories, efforts such as
\emph{deterministic builds} allow users to verify that a compiled binary
corresponds to the published source code of open-source software, a
traditionally difficult process due to sources of non-determinism in build
processes. Deterministic builds are achieved by recording the environment when
building software, then replaying the behavior of this environment in later
builds to achieve the same results~\cite{xavierdecarnedecarnavalet2014}.
While this prevents developers from inserting malicious code into the compiled
binaries (i.e., making their code public but including a different version in
the actual binary), it does not address the targeted malware threat that
\sysname aims to solve, in which the source code (or binary) for one targeted
set of users is different from the copy received by everyone else.


\subsection{Distributed ledgers}

The concept of a blockchain was first used in Bitcoin, which is designed to be
a globally consistent append-only ledger of financial
transactions~\cite{satoshinakamoto2008}.
Given our limited usage of Bitcoin, we focus for brevity only
on the properties that we require for \sysname.  

Briefly, the Bitcoin blockchain is (literally) a chain of blocks.  Each block
contains two components: a \emph{header} and a list of transactions.  In
addition to other metadata, the header stores the hash of the block
(which, in compliance with the proof-of-work consensus mechanism, must be
below some threshold in order to show that a certain amount of so-called
``hashing power'' has been expended to form the block),
the hash of the previous block (thus enabling the chain property), and the
root of the Merkle tree that consists of all transactions in the block.

On the constructive side, while the scripting language used by Bitcoin is
(intentionally) limited in its functionality, Bitcoin transactions
can nevertheless store small amounts of arbitrary data.  This makes Bitcoin
potentially useful for other applications that may require the properties of
its ledger, such as certifying the ownership and timestamp of a
document~\cite{massimobartoletti2017}.  One mechanism that allows Bitcoin to
store such data is the script opcode
\verb#OP_RETURN#,\footnote{\url{en.bitcoin.it/wiki/OP_RETURN}} which can be
used to embed up to 80 bytes of arbitrary data.

Another aspect of Bitcoin that enables additional development is the idea of
an SPV (Simplified Payment Verification)
client.
Rather than perform the
expensive verification of the digital signatures contained in Bitcoin
transactions, or the checks necessary to determine whether or not
double-spending has taken place, these clients check only that a given
transaction has made it into some block in the blockchain.  As this can be
achieved using only the root hashes stored in the block headers, such clients
can store only these headers (which are small) and verify only Merkle proofs
of inclusion obtained from ``full'' nodes (which is fast), and are thus
significantly more efficient than their full node counterparts.

On the destructive side, various attacks have been demonstrated that undermine
the security guarantees of Bitcoin.  In \emph{eclipse}
attacks~\cite{heilman-eclipse,gervais-eclipse,hijacking-bitcoin}, an adversary
exploits the topology of the Bitcoin network to interrupt, or at least delay,
the delivery of announcements of new transactions and blocks to a victim node.
More expensive ``51\%'' attacks, in which the adversary controls more than
half of the collective hashing power of the network, allow the adversary to
fork the blockchain, and it has been demonstrated~\cite{ittayeyal2014} that
such attacks can in fact be carried out with far less than 51\% of the hashing
power.

\section{Threat Model and Setting}\label{sec:threat}

In this section, we describe the actors in the ecosystem for software
distribution transparency (Section~\ref{sec:participants}), along with
the interactions between these actors (Section~\ref{sec:interactions}), and
the goals we hope to achieve in this setting (Section~\ref{sec:goals}).

\subsection{Participants}\label{sec:participants}

We consider a system with five types of actors: services, authorities,
monitors, auditors, and clients.  We describe each of these types below in the
singular, but for the correct and secure functioning of a transparency overlay
we require a distributed set of auditors and monitors, each acting
independently.

\begin{sarahlist}

\item[Service:] The service is responsible for producing actions, such as the
issuance of a software update.  In order to have these binaries authorized,
they must be sent to the authority.

\item[Authority:] The authority is responsible for publishing \emph{statements}
that declare it has received a given software binary from a service.
These statements also claim that
the authority has\dash in some form\dash published these binaries in a way
that allows them to be inspected by the monitor.  
The authority is also responsible for placing its
statements into a public \emph{audit log}, where they can be efficiently
verified by the auditor.

\item[Monitor:] The monitor is responsible for inspecting the binaries
published by the authority and performing out-of-band tests to determine
their validity (e.g., to ensure that software updates do not contain malware).

\item[Auditor:] The auditor is responsible for checking specific binaries
against the statements made by the authority that claim they are published.

\item[Client:] The client receives software updates from either the authority
or the service, along with a statement that claims the update has been
published for inspection.  It outsources all responsibility to the auditor, so
in practice the auditor can be thought of as software that sits on the client
(thus making the client and auditor the same actor, which we assume
for the rest of the paper).

\end{sarahlist}

\subsection{Interactions}\label{sec:interactions}

In terms of the interactions between these entities, one of the main benefits
of \sysname\dash as discussed in the introduction\dash is that
entities do not need to engage in prolonged multi-round interactions like
gossiping, but rather pass messages atomically to one another.  As we see in
Section~\ref{sec:security-proof}, this makes it significantly more expensive for an
adversary to present undetected split views of a log by launching man-in-the-middle
attacks.  We therefore outline only non-interactive algorithms needed to
generate messages, rather than interactive protocols, and wait to specify
the exact inputs and outputs until we present our construction in
Section~\ref{sec:design}.

\begin{sarahlist}

\item[$\authority.\commit$:] The authority runs this algorithm to commit
statements to the audit log.

\item[$\authority.\proveinclusion$:] The authority runs this algorithm to
provide a proof that a specific statement is in the audit log.

\item[$\auditor.\checkinclusion$:] The auditor runs this algorithm to check
the proof of inclusion for a specific statement.

\item[$\monitor.\getcommits$:] The monitor runs this algorithm to retrieve
relevant commitments from the audit log.

\end{sarahlist}

\subsection{Goals}\label{sec:goals}

We break the goals of the system down into security goals (denoted with an S)
and deployability goals (denoted with a D).

As discussed in the introduction, it is especially crucial in the setting of
binary transparency to consider adversaries that can perform persistent
man-in-the-middle attacks, as it is realistic that they would be able to
compromise the client's machine.  Like certificate transparency (but unlike key
transparency), we do not need to make the contents of the audit log private, as
binaries are assumed to be public information, but we do need to guarantee
privacy for the specific binaries that a client downloads, as this could reveal
that a client has a software version susceptible to malware.  Finally, even
though binaries are typically large, we need to nevertheless provide a solution
efficient enough to be deployed in practice.

Keeping these requirements in mind, we aim in all our security goals to defend
against the specified attacks in the face of malicious authorities that, in
addition to performing all the usual actions of the authority, can also perform
man-in-the-middle attacks on the auditor's network communications.  If
additional adversaries are considered we state them explicitly.

\begin{sarahlist}

\item[S1: No split views.] We should prevent split-view attacks, in
which the information contained in the audit log convinces the auditor that
the authority published a binary, and thus it is able to be inspected by
monitors, whereas in fact it is not and only appears that way in the
auditor's ``split'' view of the log.

\item[S2: Availability.] We should prevent attacks on availability, in
which the information contained in the audit log convinces the auditor that
a binary is available to be inspected by monitors, when in
fact the authority has not published it or has, after the initial publication,
lost it or intentionally taken it down.

\item[S3: Auditor privacy.] We should ensure that the specific binaries
in which the auditor is interested are not revealed to any other parties.  We
thus consider how to achieve this not only in the
face of malicious authorities, but in the case in which all parties aside from
the auditor are malicious.

\item[D1: Efficiency.] \sysname should operate as efficiently as
possible, in terms of computational, storage, and communication costs.  In
particularly, the overhead beyond the existing requirements
for a software distribution system should be minimal.

\item[D2: Minimal setup.] In addition to the computational overheads, we would
like as little effort\dash in terms of, e.g., coordination\dash to be done as
possible in order to deploy \sysname, and for it to require the minimal amount
of change to the existing system.

\end{sarahlist}

\section{Design of \sysname} \label{sec:design}

\begin{figure}[h]
\centering
\includegraphics[width=0.8\linewidth,keepaspectratio]{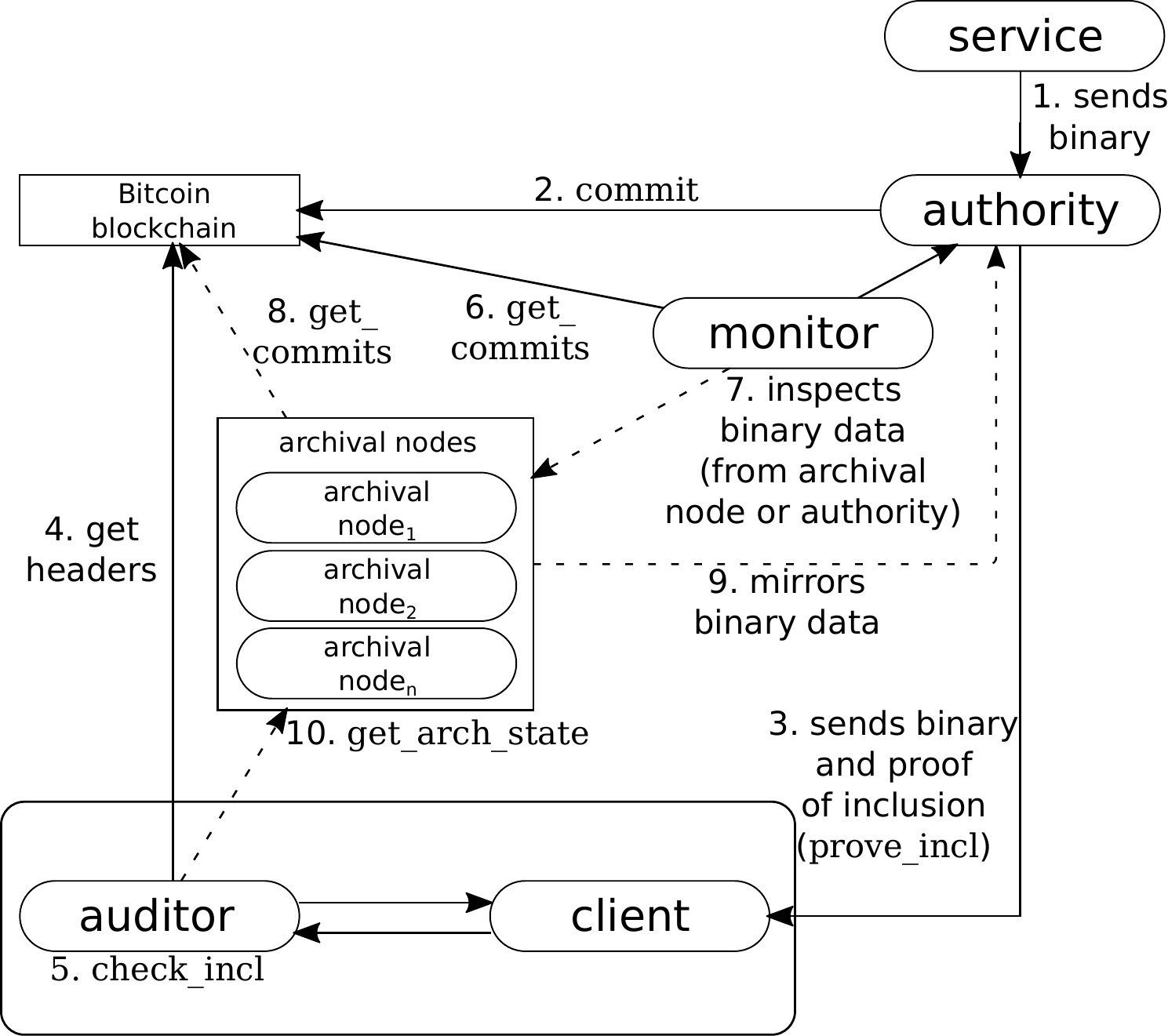}
\caption{The overall structure of \sysname. Dashed lines represent
steps that are required only if archival nodes are used.}
\label{fig:design}
\end{figure}

In this section we describe the overall design of \sysname.  An overview of
all the interactions in the system can be seen in
Figure~\ref{fig:design}.

\subsection{Setup and instantiation}\label{sec:design-setup}

\sysname and its security properties make use of a blockchain, whose primary
purpose\dash as we see in Section~\ref{sec:security-proof}\dash is to provide an
immutable ledger that prevents split-view attacks.  Because the Bitcoin
blockchain is currently the most expensive to attack, we use it here and in our
security analysis in Section~\ref{sec:security-proof}, but observe that any
blockchain could be used in its place.
An authority must initially establish a known Bitcoin address that \sysname
commitments are published with. As knowledge of the private key associated with
the Bitcoin address is required to sign transactions to spend transaction
outputs sent to the address, this acts as the root-of-trust for the authority.
This address can be an embedded value in the auditor software. An initial amount
of coins must be sent to the Bitcoin address to enable it to start making
transactions from the address.

\subsection{Logging and publishing statements}\label{sec:design-logging}

To start, the authority receives information from services; i.e., software
binaries from the developers of the relevant packages (Step~1 of
Figure~\ref{fig:design}).  As it receives such a binary, it
incorporates its hash as a leaf in a Merkle tree with root $\roothash_T$.
The root, coupled with the path down to the leaf representing the binary, thus
proves that the authority has seen the binary, so we view the root as a batched
statement attesting to the fact that the authority has seen all the binaries
represented in the tree.
Once the Merkle tree reaches some (dynamically chosen) threshold $n$ in
size, the authority runs the $\commit$ algorithm (Step~2 of
Figure~\ref{fig:design}) as follows:

\begin{sarahlist}

\item[$\commit(\roothash_T)$:] Form a Bitcoin transaction in which one of the
outputs embeds $\roothash_T$ by using \verb#OP_RETURN#. One of the inputs must
be a previous transaction output that can only be spent by the authority's
Bitcoin address (i.e. a standard Bitcoin transaction to the authority's
address). The other outputs are optional and may simply send the coins back to
the authority's address, according to the miner's fees it wants to pay. (See
Section~\ref{sec:performance} for some concrete choices.)  Sign the transaction
with the address's private key and publish to the Bitcoin blockchain and return
the raw transaction data, denoted $\tx$. \end{sarahlist}

Crucially, the $\commit$ algorithm stores only the root hash in the
transaction, meaning its size is independent of the number of statements it
represents.  Furthermore, if the blockchain is append-only\dash i.e., if double
spending is prevented\dash then the log represented by the commitments in the
blockchain is append-only as well.

\subsection{Proving inclusion}\label{sec:design-prove-inclusion}

After committing a batch of binaries to the blockchain, the authority can now
make these binaries accessible to clients.  When a client requests a software
update, the authority sends not only the relevant binary, but also an
accompanying proof of inclusion, which asserts that the binary has been placed
in the log and is thus accessible to monitors (Step~3 of
Figure~\ref{fig:design}).

\begin{figure}[h]
\centering
\includegraphics[width=0.8\linewidth,keepaspectratio]{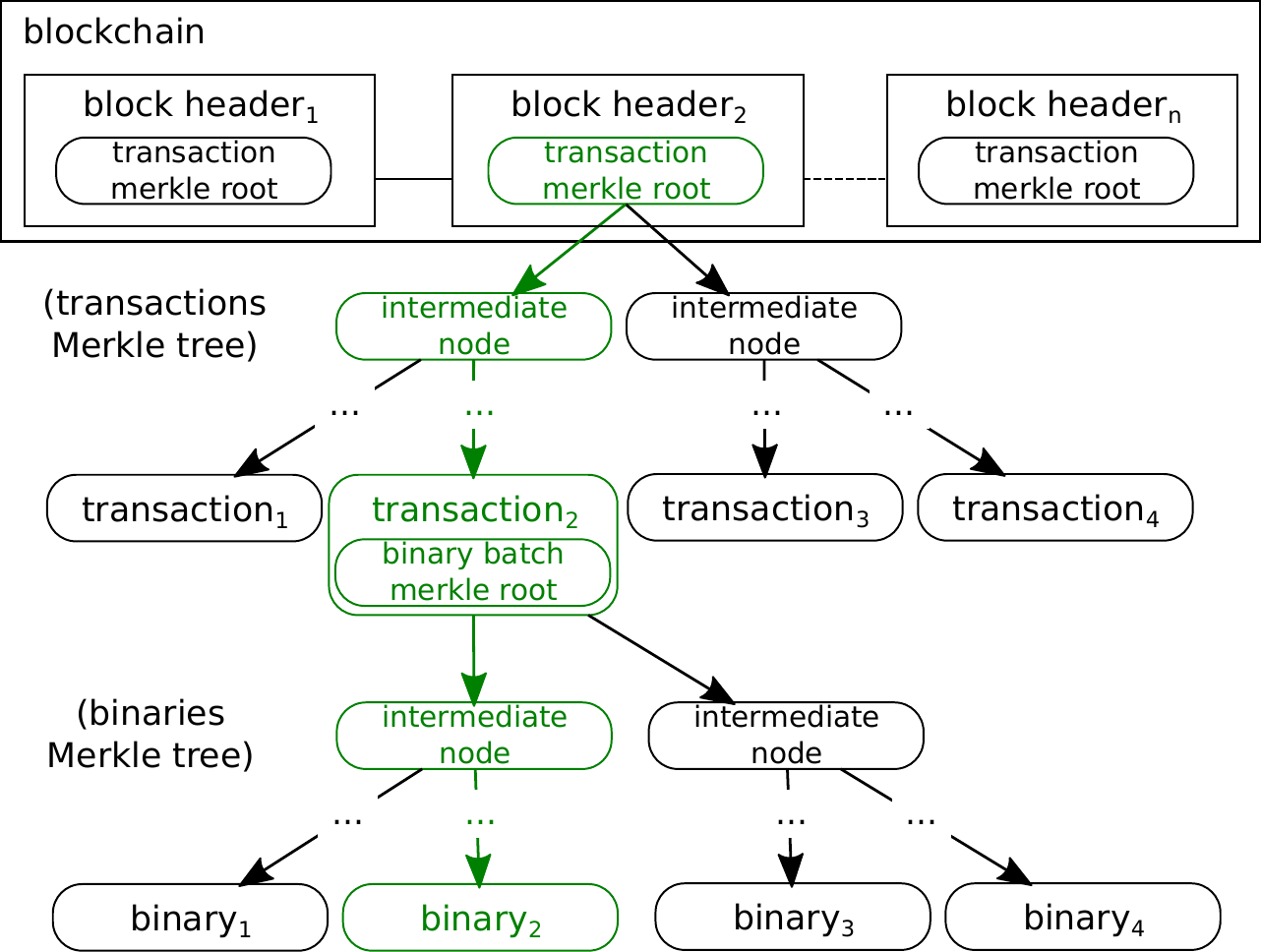}
\caption{An example of a path of hashes leading from the block's
transactions Merkle root to the hash of $\ourbinary_2$.}
\label{fig:example-proof}
\end{figure}

To generate this proof, the authority must first wait for its transaction to
be included in the blockchain (or, for improved security, for it to be embedded
$k$ blocks into the chain).  We denote the header of the block in which it was
included as $\head_B$.  The proof
then needs to convince anyone checking it of two things: (1) that the relevant
binary is included in a Merkle tree produced by the authority and (2) that the
transaction representing this Merkle tree is in the blockchain.  Thus, as
illustrated in Figure~\ref{fig:example-proof}, this means providing a path of
hashes leading from the values retrieved from the blockchain to a hash of the
statement itself.

For a given binary $\ourbinary$, the algorithm $\proveinclusion$ thus runs as
follows:

\begin{sarahlist}

\item[$\proveinclusion(\tx,\head_B,\ourbinary)$:] First, form a Merkle
proof for the inclusion of $\tx$ in the block represented by $\head_B$.  This
means forming a path from the root hash stored in $\head_B$ to the leaf
representing $\tx$; denote these intermediate hashes by $\pi_\tx$.  Second,
form a Merkle proof for the inclusion of $\ourbinary$ in the Merkle tree
represented by $\tx$ (using the hash $\roothash_T$ stored in the
\verb#OP_RETURN# output) by forming a path from $\roothash_T$ to the leaf
representing $\ourbinary$; denote these intermediate hashes by $\pi_\ourbinary$.
Return $(\head_B,\tx,\pi_\tx,\pi_\ourbinary)$.

\end{sarahlist}

\subsection{Verifying inclusion}\label{sec:design-check-inclusion}

To verify this proof, the auditor must check the Merkle proofs, and must also
check the authority's version of the block header against its own knowledge of
the Bitcoin blockchain.  This means that the auditor must first keep
up-to-date on the headers in the blockchain, which it obtains by running an
SPV client (Step~4 in Figure~\ref{fig:design}).  By running this client, the
auditor builds up a set $S = \{\head_{B_i}\}_i$ of block headers, which it can
check against the values in the proof of inclusion.  This means that, for a
binary $\ourbinary$, $\checkinclusion$ (Step~5 in Figure~\ref{fig:design})
runs as follows:

\begin{sarahlist}
\item[$\checkinclusion(S,\ourbinary,(\head_B,\tx,\pi_\tx,\pi_\ourbinary))$:]
First, check that $\head_B\in S$; output $0$ if not.  Next, extract
$\roothash_T$ from $\tx$ (using the hash stored in the \verb#OP_RETURN#
output), form $h_\ourbinary\gets H(\ourbinary)$, and check that
$\pi_\ourbinary$ forms a path from the leaf $h_\ourbinary$ to the root
$\roothash_T$.  Finally, form $h_\tx\gets H(\tx)$, and check that $\pi_\tx$
forms a path from the leaf $h_\tx$ to the root hash in $\head_B$.  If both
these checks pass then output $1$; otherwise output $0$.
\end{sarahlist}

As well as verifying the inclusion proof, the auditor must also check that the
address that the proof's transaction was sent from matches the authority's
address (i.e. one of the transaction inputs must be a previous transaction
output that can only be spent by the authority's address).

\subsection{Ensuring availability}\label{sec:design-archive}

Independently of auditors, monitors must retrieve all commitments associated
with the authority from the blockchain and mirror their binaries (Steps~6
and~7 of Figure~\ref{fig:design}).  This means $\getcommits$ runs as follows:

\begin{sarahlist}
\item[$\getcommits()$:] Retrieve all transactions in the blockchain sent
with the authority's address, and return the hashes stored in the
\verb#OP_RETURN# outputs.
\end{sarahlist}

After checking the binaries against their commitments, the monitors then
inspect them\dash to, e.g., ensure they are not
malware\dash in ways we consider outside of the scope of this paper.

While the system we have described thus far functions correctly and allows
monitors to detect if an authority has committed to a binary but not published
it, in order to make the binaries themselves available for inspection, we assume
the monitors can mirror the authority's logs.  It therefore fails to satisfy our
goal of availability in the event that the authority goes down at some point in
time.

We thus consider the case where the authority commits binaries to
the blockchain, but\dash either intentionally or because it loses the data
sometime in the future\dash does not supply the data to monitors.
While this is detectable, as monitors can see that there are
commitments in the blockchain with no data behind them, to disincentive this
behavior requires some retroactive real-world method of punishment.
More importantly, it prevents the monitor from pinpointing specific
bad actions, such as malicious binaries, and thus from identifying
potential victims of the authority's misbehavior.

Because of this, it is thus desirable to not only enable the detection of this
form of misbehavior, but in fact to prevent it from happening in the first
place.  One way to achieve this is to have auditors mirror the binary themselves
and send it to monitors before accepting it, to ensure that they have seen it
and believe it to be benign.  While this would be effective, and is arguably
practical in a setting such as Certificate Transparency (modulo concerns about
privacy) where the objects being sent are relatively small, in the setting of
software distribution\dash where the objects being sent are large binaries\dash
it is too inefficient to be considered.

Instead, we propose a new actor in the ecosystem presented in
Section~\ref{sec:threat}: archival nodes, or \emph{archivists}, that are
responsible for mirroring all data from the authority (Steps~8 and~9 in
Figure~\ref{fig:design}).  To gain the extra guarantee that the data is
available to monitors, auditors may thus use any archival nodes of which they
are aware to check their state (i.e., the most recent block header for which
they have data from the authority) and ensure that they cover the block
headers relevant to the proofs they are checking (Step~10 in
Figure~\ref{fig:design}).  This means adding the following two
algorithms to the list in Section~\ref{sec:interactions}:

\begin{sarahlist}

\item[$\archivist.\getcommits()$:] The archivist runs this algorithm to
access the commitments made by the authority, just as is done by the
monitor (using the same algorithm).

\item[$\auditor.\getarchstate()$:] The auditor (optionally) runs this
algorithm to obtain the state of any archivists of which it is aware.  This
is simply the latest block header for which the archival node has mirrored the
data behind the commitments held within.

\end{sarahlist}

Using archival nodes makes it possible to continue to pinpoint specific bad
actions in the past (e.g., the publication of malware), even if the authority
loses or stops providing this data, but we stress that their usage is optional
and affects only availability.  Essentially, archival nodes allow for a more
granular detection of the misbehavior of an authority, but do come at the
cost of requiring additional nodes to store a potentially large amount of
data.  If such granularity is not necessary, or if the system has no natural
candidates with the necessary storage requirements, then archival nodes do
not need to be used and the system still remains secure.  In
Section~\ref{sec:debian} we explore the role of the archival nodes in the
Debian ecosystem and discover that, while the storage costs are indeed
expensive, there is already at least one entity playing this role.

\section{Evaluation}\label{sec:evaluation}

In this section, we evaluate \sysname in terms of how well it meets the
security goals (Section~\ref{sec:security-proof}) and deployability goals
(Section~\ref{sec:deploy-proof}) specified in our threat model in
Section~\ref{sec:goals}.  We also compare it with respect to previous
solutions in Section~\ref{sec:compare-others}, and argue that it is
the only system to achieve all our goals.

\subsection{Security goals}\label{sec:security-proof}

\subsubsection{No split views (S1)}

In order to prevent split views, we rely on the security of the Bitcoin
blockchain and its associated proof-of-work-based consensus mechanism.
If every party has the same view of the blockchain, then split views of the log
are impossible, as there is a unique commitment to the state of the log at any
given point in time. The ability to prevent split views therefore reduces to the
ability to carry out attacks on the Bitcoin blockchain.

If, for whatever reason, the adversary cannot carry out an eclipse attack,
then it can perform a split-view attack only if it can fork the Bitcoin
blockchain. This na{\"i}vely requires it to control 51\% of the network's
mining power, which we estimate would cost roughly 2043M USD in electricity
and hardware costs as of December 2017 (see Appendix~\ref{sec:appendix} for
the analysis).  Regardless of the exact number, it is generally agreed that
carrying out such an attack is prohibitively expensive.

If an eclipse attack is possible, due to the adversary's MitM capability,
the adversary can ``pause'' the auditor at a
block height representing some previous state of the log, and can prevent the
auditor from hearing about new blocks past this height.  It is then free to mine
blocks at its own pace, and so performing a split-view attack would be
significantly cheaper. As a key distinguishing property of \sysname's threat
model is that split-view attacks should be prevented even in the face of
an adversary that can carry out such attacks, it is important
to consider the nuances and costs of this attack, especially as
we are not aware of any previous literature considering the costs of
eclipse attacks on Bitcoin nodes.

The cost of performing an eclipse attack depends on how much time the adversary
has to perform a split-view attack, as the hash rate depends on the number of
mining rigs available. As a rough estimate (see Appendix~\ref{sec:appendix}
for calculations), if auditors consider a Bitcoin transaction to be confirmed
after 6 blocks (the
standard for most Bitcoin wallets), then as of December 2017 the attack would
cost 8.3M USD if the adversary wants to perform the attack within a week.
This would mean, however, that the auditor would receive a new block only
every 1.4 days, which would be detectable as an eclipse attack. If
auditors conservatively require that new blocks arrive in intervals of up to
three hours before
assuming that they are the victim of an eclipse attack, then as of December 2017
an attack would cost roughly 91.8M USD.

\subsubsection{Availability (S2)}

While the decentralized (and thus fully replicated) nature of the blockchain
can guarantee availability, it guarantees these properties only with respect
to the commitments to statements made by the authority, rather than with
respect to the statements\dash and thus the binaries\dash themselves.  As
discussed in Section~\ref{sec:design-archive}, the use
of the blockchain thus does not guarantee that binaries are actually available
for inspection, or will continue to be into the future.

Even just using monitors, \sysname can already detect that an authority
committed a statement without making the statement data (i.e., the actual
binaries) available.  Using the archival nodes introduced in
Section~\ref{sec:design-archive}, we can achieve a stronger notion of
availability\dash in which as long as the binaries have been published at some
point they can be retrieved indefinitely into the future\dash as long as
these nodes are honest about
whether or not they have mirrored the relevant data.

In binary transparency, many ISPs and hosting providers already
provide their customers local mirrors of Debian repositories. We therefore
envision that ISPs can act as archival nodes on behalf of their hosting
clients, which creates a decentralized network of archival nodes.  We
elaborate on the overheads required to do so in Sections~\ref{sec:performance}
and~\ref{sec:debian}.

\subsubsection{Auditor privacy (S3)}

Recall from Section~\ref{sec:interactions} that one of the goals of \sysname was
to avoid prolonged interactions and engage only in the atomic exchange of
messages.  In particular, the auditor receives pre-formed proofs of inclusion
from the authority (as opposed to having to request them for specific
binaries, as they would in all certificate and key transparency systems),
retrieves commitments directly from the blockchain, does not engage in any form
of gossip with monitors, and receives the latest block hash from archival nodes
without providing any input of its own.  We thus achieve privacy by design, as
at no point does the auditor reveal the statements in which it is
interested to any other party.

One particular point to highlight is that \sysname achieves auditor privacy
despite the fact that auditors run SPV clients, which are known to potentially
introduce privacy issues due to the use of Bloom filtering and the reliance on
full nodes.  This is because the proofs of inclusion contain both the raw
transaction data and the block header, so the auditor does not need query
a full node for the inclusion of the transaction and can instead verify it
itself (and, as a bonus, saves the bandwidth costs of doing so).

\subsection{Deployability goals}\label{sec:deploy-proof}

\subsubsection{Efficiency (D1)}

Table~\ref{tab:complexity} summarizes the computational complexity of each of
the operations required to run \sysname, and Table~\ref{tab:size-complexity}
summarizes the size complexity (which in turn informs the bandwidth
requirements, as we explore further in Section~\ref{sec:performance}).

\begin{table}
\centering
\begin{tabular}{lc}
\toprule
Operation & Time complexity\\
\midrule
$\commit$ & $O(n_S)$\\
$\proveinclusion$ (one-time) & $O(\log(n_T))$\\
$\proveinclusion$ (per statement) & $O(\log(n_S))$\\
$\checkinclusion$ & $O(\log(n_S)+\log(n_T))$\\
\bottomrule
\end{tabular}
\tcaption{Asymptotic computational costs for the operations of \sysname,
where $n_S$ is the number of statements in a batch and $n_T$ is the number
of transactions in a block.}
\label{tab:complexity}
\end{table}

\begin{table}
\centering
\begin{tabular}{lc}
\toprule
Object & Size Complexity\\
\midrule
Inclusion proof & $O(\log(n_S)+\log(n_T))$\\
Log commitment ($\tx$) & $O(1)$\\
Archival node data overhead & $O(n_S)$\\
\bottomrule
\end{tabular}
\tcaption{Asymptotic storage costs for the objects in \sysname, where $n_S$ is
the number of statements in a batch and $n_T$ is the number of transactions in
a block.}
\label{tab:size-complexity}
\end{table}

As we will see in Sections~\ref{sec:performance} and~\ref{sec:debian}, in real
deployments of \sysname there are already significant storage costs for the
authority and archival nodes, as they must store the full set of binaries.  It
therefore does not impose a significant additional burden to have them perform
relatively inefficient (i.e., linear in $n$) operations or store relatively
inefficient objects.  As for the end-user devices on which the auditor is run,
we impose a relatively minimal performance overhead (with everything
logarithmic in $n_S$ and/or $n_T$), and confirm this in
Section~\ref{sec:auditor-performance}.

\subsubsection{Minimal setup (D2)}

In terms of coordination, the only setup requirement in \sysname is the role of
the archival nodes, as the rest is just a matter of adding software.  As we will
see in Section~\ref{sec:debian} when we look at Debian, in some settings there
are already natural candidates for these actors, but if these actors are not
interested in the guarantees of \sysname then we can still deploy it without
requiring the existing actors to change their behavior.  More importantly,
there are no trust requirements placed on these nodes to prevent log
equivocation: even if archival nodes misbehave, monitors can still
individually detect misbehavior by an authority that publishes commitments
but not the underlying data.  This is in stark contrast to previous solutions
that require the initial establishment of a semi-trusted set of nodes.

\subsection{Comparison with existing solutions}\label{sec:compare-others}

To fully pinpoint both the benefits and tradeoffs of \sysname, we compare it
with several known systems designed to provide transparency.  In particular,
we consider the tradeoffs as compared to Certificate Transparency (CT),
Collective Signing (CoSi)~\cite{ewasyta2015}, CONIKS~\cite{coniks}, and
Bitcoin.  We summarize these tradeoffs in Table~\ref{tab:comparisons}.

\begin{table*}
\centering
\begin{tabular}{lcccccc}
\toprule
& \multicolumn{3}{c}{Security goals (S1-S3)} &
    \multicolumn{3}{c}{Deployability goals (D1-D2)}\\
\cmidrule(lr){2-4} \cmidrule(lr){5-7}
& Split views & Availability &
    Auditor privacy & Efficiency (cost) & Efficiency (size) & Minimal setup\\
\midrule
CT & detect & no* & no & $\log(n)$ & 1 & no\\
CoSi & prevent & yes* & yes & 1 & 1 & no\\
CONIKS  & detect & no & no & $\log(n)$ & 1 & no\\
Bitcoin & prevent & yes & yes & $n$ & $n$ & yes\\
\sysname & prevent & yes & yes & $\log(n)$ & $b$ & yes\\
\bottomrule
\end{tabular}
\tcaption{A comparison between existing solutions and \sysname in terms of
the five goals presented in Section~\ref{sec:goals}.  For efficiency, we
measure the asymptotic costs for the auditor in terms of both the
computations it must perform (`cost') and the data it must store (`size').  We
use $n$ to denote the number of statements and $b$ to denote the number (but
not size) of blocks in the Bitcoin blockchain.  For CoSi, availability is not
a explicit requirement, but can be satisfied as long as at least one witness
retains the data, and for CT it is not satisfied by the basic design
but could be if auditors and monitors gossiped about certificates.}
\label{tab:comparisons}
\end{table*}

Looking at Table~\ref{tab:comparisons}, we first mention that the efficiency
numbers for CoSi are somewhat misleading, as there is no global log and thus
no notion of checking inclusion in the log; this is why we list the efficiency
costs as constant.  In fact, only Bitcoin and \sysname ensure a globally consistent
ledger, as certificates are stored in a distributed set of logs in CT and
CONIKS and there is no proposed method for achieving consensus amongst them.

Arguably the main benefit of both CT and CONIKS is their efficiency, as the
auditor is required to store
only a single hash.  The tradeoff, however, is that they cannot prevent the
authority from launching a split-view attack, but instead rely on gossiping
mechanisms to detect such misbehavior after the fact.  As discussed in the
introduction, this is problematic in a setting\dash like binary
transparency\dash in which adversaries can launch
persistent man-in-the-middle attacks.  These systems also do not achieve
robust privacy for the auditor, as it must periodically reveal information
to the authority (or the monitor) about the objects in which it is interested.

The other main tradeoff we observe is, perhaps unsurprisingly, between
efficiency and setup costs.  The first three systems all require the
establishment of some initial set of distributed entities\dash in the case of
CT, log servers are essentially authorized by Google, in the case of CONIKS,
identity providers are chosen by users and listed in a PKI, and in the case
of CoSi, witnesses must form a Sybil-free set\dash that are trusted to some
extent (if not individually, then as a group).  We require no
such setup, which means \sysname is much more easily integrated into existing
systems.

In contrast, in both Bitcoin
and \sysname, the blockchain is maintained by a decentralized network and is
not subject to intervention by central authorities.  While \sysname mitigates
the inefficiency of Bitcoin, it still requires the auditor to store some
information from all the block headers.  We show in the next two
sections that \sysname is nevertheless efficient enough to be practical, but
leave it
as an interesting open problem to investigate to which extent
these tradeoffs between efficiency and decentralization are inherent.

\section{Implementation and Performance}\label{sec:impl}

To test \sysname and analyze its performance, we have implemented and provided
benchmarks for a prototype Python module and toolset that developers can use. We
have released the implementation as an open-source project.\footnote{https://github.com/musalbas/contour}

\subsection{Implementation details}

The implementation consists of roughly 1000 lines of Python code, and provides a
set of developer APIs and corresponding command-line tools.  We used SHA-256 as
the hashing algorithm to build Merkle trees, and modified versions (for Bitcoin
compatibility) of an existing Merkle tree implementation
(\url{https://github.com/jvsteiner/merkletree}) and a Python-based Bitcoin
library \texttt{pycoinnet} (\url{https://github.com/richardkiss/pycoinnet/}) in
order to develop our Merkle tree and SPV client, respectively.

\begin{sarahlist}

\item[Authority:]
We provide API calls for $\authority.\commit$, which commits batches of
statements to the Bitcoin blockchain, and $\authority.\proveinclusion$,
which allows it to generate inclusion proofs for individual statements.

\item[Auditor:]
We provide an API call for $\auditor.\checkinclusion$, which allows end-user
software to verify proofs of inclusion.  We also provide an
$\auditor.\sync$ call that uses the Bitcoin SPV protocol to download and verify
all the block hashes in the Bitcoin blockchain, so that inclusion proofs can be
efficiently verified independently of third parties. (This call needs to be
run only once.)

\item[Monitor:]
We provide an API call for $\monitor.\getcommits$, which gets all the
statement batches associated with a specific authority.  Monitors can then
use these commitments to check the validity of the statement data (which they
can retrieve
from the authority or an archival node using a web server), and do whatever
manual inspection is necessary; we consider this functionality outside of the
scope of this paper.

\item[Archival node:]
The archival node API can be used to operate an archival node, by
specifying the authority's Bitcoin address and web address where statement data
is published. The archival state and mirrored statement data is stored as flat
files on disk, allowing the archival node to provide access to auditors and
monitors by running a web server. By accessing the archival state via a HTTPS
server, auditors can securely authenticate the state of the archival node using
public-key cryptography.

\end{sarahlist}

\subsection{Performance}\label{sec:performance}

To evaluate the performance of our implementation, we tested all the
operations listed above on a laptop with an Intel Core i5 2.60\;GHz CPU and
\SI{12}{\giga\byte} of RAM, that was connected to a WiFi network with an
Internet connection of 5\;Mbit/s.  We also assume that a batch to be
committed contains 1 million
statements, although as was seen in Table~\ref{tab:complexity}\dash and
will be confirmed later on in Figure~\ref{fig:checkinclusion}\dash these
numbers scale as expected (either logarithmically or linearly), so it is easy to
extrapolate the results for other batch sizes given the ones we present here.

We consider the complexity of these operations in terms of their
computational, storage, and bandwidth requirements.  A summary of our
timing benchmarks can be found in Table~\ref{tab:benchmarks}, and our
bandwidth requirements are in Table~\ref{tab:bandwidth}.

\begin{table}
\centering
\begin{tabular}{lS[table-format=3.1,detect-weight]S[table-format=2.2]}
\toprule
Operation & {Time (\si{\microsecond})} & {$\sigma$ (\si{\microsecond})}\\
\midrule
$\commit$ & {\bfseries 5.93 (s)} & {\bfseries 0.297 (s)} \\
$\proveinclusion$ (one-time) & 8.5 & 5.4 \\
$\proveinclusion$ (per statement) & 12 & 6.4 \\
$\checkinclusion$ & 224 & 62.14 \\
\bottomrule
\end{tabular}
\tcaption{Average time of individual operations, and standard deviation
$\sigma$, when the batch size is 1M.  The timings for
$\commit$ were averaged over 20 runs, and for $\proveinclusion$ and $\checkinclusion$ over 1M runs.  The timings for $\commit$ are in bold to emphasize that they are in seconds, not microseconds.}
\label{tab:benchmarks}
\end{table}

\begin{table}
\centering
\begin{tabular}{lS[table-format=3.1]}
\toprule
Operation & {Bandwidth}\\
\midrule
$\authority.\commit$ (using APIs) & \SI{1}{\mega\byte} \\
$\authority.\commit$ (one-time setup for full node) & \SI{126}{\giga\byte} \\
$\authority.\commit$ (using full node) & \SI{235}{\byte} \\
$\auditor.\sync$ & \SI{37.4}{\mega\byte} \\
$\auditor.\proveinclusion$ & \SI{1.3}{\kilo\byte} \\
\bottomrule
\end{tabular}
\tcaption{The bandwidth cost of operations, when the batch size is 1M.
The cost of $\authority.\commit$ depends on whether or not the
authority is running a full Bitcoin node or relying on third party APIs.  For
running a full node, there is a one-time setup cost to synchronize the
blockchain.}
\label{tab:bandwidth}
\end{table}

\subsubsection{Number of transactions per block}\label{sec:full-block}

The overhead of both generating and verifying a proof of inclusion is
dependent on the number of transactions in a Bitcoin block.  To capture the
worst-case scenario, we consider the maximum number of transactions that can
fit into a block.  Currently, the
Bitcoin block size limit is \SI{1}{\mega\byte}, up to 97 bytes of which
is non-transaction data.\footnote{\url{https://en.bitcoin.it/wiki/Block}}
The minimum transaction size is 166
bytes,\footnote{\url{https://en.bitcoin.it/wiki/Maximum_transaction_rate}}
so the upper
bound on the number of transactions in a given block is 6,023.  While this
is far higher than the number of transactions that Bitcoin blocks currently
contain,\footnote{\url{https://blockchain.info/charts/n-transactions-per-block}}
we nevertheless use it as a
worst-case cost and an acknowledgment that Bitcoin is evolving and blocks may
grow in the future.

\subsubsection{Authority overheads}\label{sec:authority-performance}

To run $\commit$ and $\proveinclusion$, an authority must have access to the
full blocks in the Bitcoin blockchain, as well as the ability to broadcast
transactions to the network.  Rather than achieve these by running the authority
as a full node, our implementation uses external blockchain APIs supplied by
\url{blockchain.info} and \url{blockcypher.com}.  This decision was based on
the improved efficiency and ease of development for prototyping, but it
does not affect the security of the system: authorities do not need to
validate the blockchain, as invalid blocks from a dishonest external
API simply result in invalid inclusion proofs that are rejected by the
auditor.

To run $\commit$, an authority must first build the Merkle tree containing its
statements.  Sampled over 20 runs, the average time to build a Merkle
tree for 1M statements was \SI{5.9}{\second} ($\sigma = $ \SI{0.29}{\second}).
After building the tree, an authority next embeds its root hash (which is 32
bytes) into an \verb#OP_RETURN# Bitcoin
transaction to broadcast to the network.  Sampled over 1,000 runs, the
average time to generate this transaction\dash in the standard case of one
input and two outputs, one for \verb#OP_RETURN# and one for the authority's
change\dash was \SI{0.03}{\second} ($\sigma = $\SI{0.007}{\second}).  The
average total time to run $\commit$ was thus \SI{5.93}{\second}, as seen in
Table~\ref{tab:benchmarks}, and it resulted in 235 bytes (the size of the
transaction) being broadcast to the network.

Next, to run $\proveinclusion$, the authority proceeds in two phases: first
constructing the Merkle proof for its transaction within the block where it
eventually appears, and next constructing the Merkle proof for each statement
represented in a transaction.  The time for the first phase, averaged over 1M
runs and for a block with 6,023 transactions (our upper bound from
Section~\ref{sec:full-block}), was \SI{8.5}{\microsecond}.  This is denoted
``one-time'' in Table~\ref{tab:benchmarks} as it is done only once per batch.
The time for the second phase, averaged over 1M runs, was \SI{12}{\microsecond}
for each individual statement (thus denoted ``per statement'' in
Table~\ref{tab:benchmarks}).  Generating inclusion proofs for all the statements
in the batch would thus take around \SI{12}{\second}.  In terms of bandwidth and
storage, a block up to \SI{1}{\mega\byte} in size needs to be downloaded in
order to generate the inclusion proof from the block's transaction Merkle tree.
In terms of the memory costs, the size of the Merkle tree for 1M leaves in
memory is \SI{649}{\mega\byte}.

Additionally, in order to ensure that its transaction makes it into a block
quickly, the authority may want to pay a fee.  The recommended rate as of December
5 2017 is 154 satoshis/byte (\url{https://bitcoinfees.info}), so
for a 235-byte transaction the authority can expect to pay 36,190 satoshis.
As of December 5 2017, this is roughly 4.21 USD.  We stress, however, that the
Bitcoin price is notoriously volatile (for example, the same transaction would
have cost only 0.28 USD at the beginning of 2017), so this and all other costs
stated in fiat currency should be taken with a grain of salt.

\subsubsection{Auditor overheads}\label{sec:auditor-performance}

For the auditor, we considered two costs: the initial cost to retrieve the
necessary header data ($\sync$), and the cost to verify an inclusion
proof ($\checkinclusion$).  We do not provide benchmarks for the
$\auditor.\getarchstate$ call, as this is a simple web request that returns a
single $32$-byte hash.

To run $\sync$, auditors use the Bitcoin SPV protocol to download and verify
the headers of each block, which are 80 bytes each. As of
December 5 2017, there are 497,723 valid mined blocks, which equates to
\SI{39.8}{\mega\byte} of block headers.  Once downloaded, however, the
auditor needs to keep only the 32-byte block hash, so only
\SI{15.9}{\mega\byte} of data needs to be stored on
disk.  Going forward, the Bitcoin network generates approximately 144 blocks
per day, so the amount of downloaded data will be \SI{11.5}{\kilo\byte}
daily, and the amount of stored data will increase by \SI{4.6}{\kilo\byte} daily.

To verify the validity of the block headers in the chain, the client must
perform one SHA-256 hash per block header; averaged over five runs, it took us
116 seconds for the Python SPV client to download and verify all the
block headers.  This initial bootstrapping process needs to be
performed only once per auditor.

To run $\checkinclusion$, we again use our upper bound from
Section~\ref{sec:full-block} and assume every block contains 6,023
transactions.  This means the inclusion proof contains: (1) an 80-byte block
header; (2) the raw transaction data, which is 235 bytes; (3) a Merkle proof
for the transaction, which consists of $\log(6023)-1$ 32-byte hashes (the
root hash is already provided in the block header); and (4)
a Merkle proof for the statement, which consists of $\log(1000000)-2$
32-byte hashes (the root hash is already provided in the transaction data, and
the auditor computes the statement hash itself).  The total bandwidth cost is
therefore around 1275 bytes.  Averaged over 1M runs, the time for the
auditor to verify the inclusion proof was \SI{224}{\microsecond} ($\sigma = $
\SI{62.14}{\microsecond}).

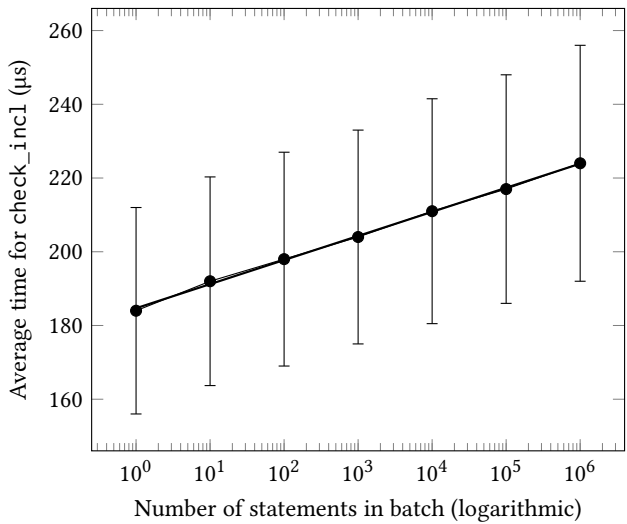
\begin{figure}
\resizebox {\linewidth} {!} {
\begin{tikzpicture}
\begin{axis}[
xmin=0,
xmode=log,
xlabel={Number of statements in batch (logarithmic)},
ylabel={Average time for $\checkinclusion$ (\si{\microsecond})},
]
\addplot [error bars/error bar style={black}, mark options={black}, mark=*] plot [
         error bars/.cd,
         y dir=both,
         y explicit,
         x dir=both,
         x explicit,
       ] table[x=X, y=Y, y error=errorY] {inclusion-proof-table.dat};
\addplot [thick] table[
y={create col/linear regression={y=Y}}
] 
{inclusion-proof-table.dat};
\end{axis}
\end{tikzpicture}
}
\vspace{-2em}
\caption{The time to verify an inclusion proof with
varying batch sizes, averaged over 100K runs.}
\label{fig:checkinclusion}
\end{figure}

To confirm that the time to run $\checkinclusion$ scales logarithmically with
the number of statements in the batch, we also ran it for varying numbers of
statements.  The results are in Figure~\ref{fig:checkinclusion}.

\subsubsection{Monitor overheads}

Monitors must run a Bitcoin full node in order to get a complete uncensored
view of the blockchain.
As of December 2017, running a full node requires \SI{145}{\giga\byte}
of free disk space, increasing by up to \SI{144}{\mega\byte} daily.  It took us
around three days to fully bootstrap a full node and verify all the
blocks, although this operation needs to be performed only once per
monitor.

\subsubsection{Archival nodes overheads}\label{sec:archive-performance}

Like monitors, archival nodes need to run a Bitcoin full node.  Additionally,
archival nodes must download and store all the data from the
authority.  The costs here are entirely dependent on the number and size of
the statements; we examine the costs for Debian in Section~\ref{sec:debian}.

In order for archival nodes to know which statement data to download from
authorities to independently rebuild the Merkle tree roots committed in Bitcoin
transactions and check that they match with the data provided, authorities must
point the archival nodes to the location of the data.  Again, this is
dependent on the mechanism that the authority uses to make the data available.

As in Debian, however, archives use statements that represent files.  We
may therefore expect that, in addition to a Merkle tree, authorities would use
metadata files to link each leaf in the tree to a file on the server that
archival nodes then mirror; this would be particularly useful in a setting\dash
like Debian\dash where it would be undesirable to reorganize files that are
already stored.  The metadata file would consist of a mapping of 32-byte hashes
to filenames.  The average Debian package filename is 60 bytes, so including
such a metadata file would introduce an average storage overhead, for both
authorities and archival nodes, of 92 bytes per statement.

\section{Use Case: Debian}\label{sec:debian}

To demonstrate how \sysname can be used on a real system, we prototyped it for
auditing software binaries in the Debian software repository.
Our results show that \sysname provides a way to add transparency
to this repository without major changes to the existing infrastructure and with
minimal overheads.  It could be deployed on top of the Debian
ecosystem today, without any participant who did not want to opt in having to
change their behavior.

We begin with an overview of how Debian currently works, and then go on to
explain how existing actors in the ecosystem could play the roles necessary
for \sysname, along with the overheads.

\subsection{Software distribution architecture}

Debian is a popular Linux distribution used by over 32\% of websites that run
Linux.\footnote{\url{https://w3techs.com/technologies/details/os-linux/all/all}}
%
Software packages are installed and updated on Debian machines using the
\verb#apt# command-line program. The Debian software repository contains
\verb#Release# files for various versions of Debian, which are updated every
time any package in the repository is updated. Each \verb#Release# file contains
a checksum for a \verb#Packages# file, which contains a list of available
software packages and their associated checksums for integrity checking.

Software packages are downloaded as \verb#.deb# archives which provide the
compiled binaries and scripts required to install a package on a system. These
files are
hosted in directories on HTTP mirrors, of which hundreds exist around
the world.\footnote{\url{https://www.debian.org/mirror/list}}

To cryptographically authenticate software packages, Debian has a set of tools
called \verb#apt-secure#. Debian installations come with a built-in set of PGP
keys~\cite{simongarfinkel1996} that are used as trusted keys for validating
software packages. Alongside the \verb#Release# files in the repository, there
are \verb#Release.gpg# files that contain PGP signatures of the \verb#Release#
files under trusted PGP keys.\footnote{\url{https://wiki.debian.org/SecureApt}}

Through the single signature of a \verb#Release# file, \verb#apt# can validate
that individual \verb#.deb# packages were authorised by a trusted PGP key by
checking that the checksums of packages are included in the \verb#Packages# file
whose checksum is included in the root \verb#Release# file. This of course
creates a central point of failure, as the owner of the signing key can serve
individual users targeted \verb#Release# files\dash for example, if coerced to
do so by law enforcement\dash that link to malicious packages.

\subsection{Authority}

In the case of Debian software distribution, the most natural operators for
a \sysname authority are the maintainers of the software repository.
Specifically, the \sysname authority would be the owner of the PGP
key, as only this entity has the power to modify the software repository.
Importantly, it is also possible for third parties to act as
\sysname authorities by proxy and commit binaries to the log on behalf of the
maintainers of the Debian software repository. As committed binaries are
transparent, the third party is not trusted any more than the maintainers of the
Debian software repository would be, as any rogue additions to the log would
still be detectable. This means it would be possible to deploy \sysname today
without any intervention or permission from the Debian project itself.

To initiate the system as an authority, all the existing software packages
would first need to be committed; i.e., the authority would need to commit to
the current state of the repository.  To measure the overhead needed for this
step, we extracted the software package metadata for all processor
architectures and releases of Debian from the Debian FTP archive
(\url{https://www.debian.org/mirror/ftpmirror}) over a one-week period from
January 20-27 2017. At the beginning of this period there were 976,214 unique
software binaries available for download from the Debian software repositories,
constituting \SI{1.7}{\tera\byte} of data, and by the end there were 980,469.

As discussed above, the Debian package metadata already contains a SHA-256
hash for every packages, so we needed only to build a Merkle tree from these
hashes (rather than compute them ourselves first), to then commit on the
blockchain. This took approximately 6 seconds (which is in line with our
benchmarks in Table~\ref{tab:benchmarks} for 1M statements).

Going forward, the authority must commit batches of new and updated binaries to
the log. The Debian FTP archives are updated four times a day, which means four
batches to commit to the log per day. Recall from
Section~\ref{sec:authority-performance} that committing one transaction to the
blockchain currently costs roughly 4.21 USD in fees, so this would cost 16.84
USD per day (although, as mentioned in Section~\ref{sec:impl}, Bitcoin prices
are notoriously volatile). This is a relatively low price to pay for a system
that costs over 91M USD to attack (Section~\ref{sec:security-proof}).

As the archive was updated, we kept track of the package hashes being added and
created a new batch for each update.  The average batch size was 1,040
packages, and the average time to build a Merkle tree for the batch was
0.0052 seconds.

As discussed in Section~\ref{sec:archive-performance}, we can also enable
archival nodes to rebuild Merkle trees with minimal changes to the existing
Debian archive infrastructure.  This requires creating and storing only an
additional \SI{84}{\kilo\byte} metadata file per batch, and an initial
\SI{79}{\mega\byte} metadata file. These metadata files consist of a mapping of
hashes of software packages to their filenames in the Debian archives.

Finally, the proof of inclusion of each software package would need to be stored
alongside each software package (\verb#.deb#) file as metadata to be downloaded
by Debian machines.  At 980K software packages, this would require a maximum of
\SI{1.3}{\kilo\byte} of extra storage per package, or \SI{1.3}{\giga\byte} of
extra storage to store the proofs of inclusion for all packages.  Given the
current storage requirements of (at least) \SI{1.7}{\tera\byte}, this is only
a $0.07\%$ overhead.

\subsection{Auditors}

On the end-user side, the \verb#apt# program would need to be modified to
integrate the $\auditor.\checkinclusion$ and $\auditor.\sync$ calls, as
implemented and analyzed in Section~\ref{sec:impl}.  This would ensure that
downloaded packages are in the log before being installed.

In terms of overhead for end-user Debian machines, as discussed above this
would require an extra \SI{1.3}{\kilo\byte} of bandwidth per package
downloaded or updated.  Given that the average package size is
\SI{1337}{\kilo\byte}, the average overhead is 0.1\% per package.  We
stress that this is a bandwidth requirement only, as once the proofs of
inclusion are verified they do not need to be stored on the client's machine.

On a freshly installed Debian 8.8 system there are 520 packages installed by
default, with a total \verb#.deb# archive size of \SI{190}{\mega\byte}.
Verifying that each of these are in the log would require an extra
\SI{698.1}{\kilo\byte} of bandwidth, and would take under two minutes.

\subsection{Monitors}

Debian's reproducible builds project allows any interested parties to verify
that binaries published in the software repositories are compiled from a given
source code.\footnote{\url{https://wiki.debian.org/ReproducibleBuilds}} There
are no specific parties assigned to the role of monitoring builds to see if
they can be built from the source code.
Similarly in \sysname, any parties vested in the security of Debian may act as a
monitor.  Aside from end users, we anticipate that large organizations supplying
critical infrastructure using Debian, national CERTs, and NGOs such as the
Electronic Frontier Foundation would have an interest in monitoring the
log.

Generally, any party that wants extra guarantees about the software updates
they are installing\dash e.g., in order to be sure that the updates that have been
pushed to their machines are the same as those that have been pushed to other
machines\dash should run a monitor. For example, if a party running Debian
receives $\mathsf{update}_1$ and $\mathsf{update}_3$ on their machine for
some software package, but the log contains $\mathsf{update}_1$,
$\mathsf{update}_2$, and $\mathsf{update}_3$, then this raises a red flag as to
why they did not receive $\mathsf{update}_2$.  In particular,
$\mathsf{update}_2$ may be a malicious update targeted to specific machines,
and the party can check to see if the contents of $\mathsf{update}_2$ have
been made available by the authority.  If they have not, then the authority
is considered to be misbehaving.  Optionally, honest archival nodes would
prevent auditors from accepting the update altogether.

\subsection{Archival nodes}

There are 269 Debian mirrors hosting the full \SI{1.7}{\tera\byte}
archive, and we view these servers as the most natural candidates for
operating archival nodes.  The difference between a mirror and an archival
node is that to fully satisfy availability an archival
node should not delete any packages (even when packages are updated and
removed), in order to enable monitors to examine obsolete packages.

In terms of overhead for archival nodes, this means storing the initial
\SI{1.7}{\tera\byte}, and then an additional average of \SI{11}{\giga\byte}
per day, or \SI{4}{\tera\byte} per year. This is by far the highest overhead
incurred by our system, and we expect that only a small number of mirrors
would have the storage capacity to run an archival node.  We stress that the
use of archival nodes is optional and serves only to boost availability (as
opposed to being required for integrity); moreover, there is currently at
least one mirror hosting all historical Debian packages, so effectively
already acting as an archival node\footnote{\url{snapshot.debian.org/}}.

\subsection{Summary}

In summary, \sysname could be deployed on top of the existing system for
Debian software distribution with minimal changes to the existing
infrastructure.  In terms of operating costs, the biggest overhead required to
enable \sysname is the extra storage space
required for archival nodes (and again, this cost is optional).  All other
costs are minimal, with only a 0.07\%
storage overhead required for the authority, and a 0.1\% bandwidth overhead
for the end user.  The computational costs for these users are minimal as
well.

One distinguishing feature of \sysname is that no
existing parties in the Debian infrastructure are required to participate if
they do not want to, and as discussed earlier the security assumptions of the
system would remain the same even if a third party acted as an authority.
This places \sysname in contrast to existing proposals for transparency
(including some of the ones presented in Section~\ref{sec:compare-others}), as
they require the initial setup of some Sybil-free set of nodes.  In contexts
such as the distribution of Debian software packages, this assumption\dash
and the security implications if it is violated\dash presents
a significant obstacle to deployability, and avoiding this obstacle was one of
our main goals in designing \sysname.

\section{Discussion and Extensions}\label{sec:discussion}

\paragraphn{Selective disclosure.} When releasing software updates that patch
critical security vulnerabilities, some software vendors may prefer not to
reveal to potential attackers that, in the window of time in which a commitment
has not yet been included in the blockchain, they can take advantage of victims
with this vulnerable software installed.  In such a case, \sysname accounts for
this by allowing the authority to commit to a batch of binaries visibly on the
blockchain, but delay the publication of the binaries themselves until the
commitment is sufficiently deep in the blockchain.
\\~\\
\paragraphn{Generalized transparency.} Although we have designed \sysname for
the specific application of binary transparency, the system is general enough to
be applied to other applications requiring transparency.  With the tradeoffs
discussed in Section~\ref{sec:compare-others}, it can even be applied to the
setting of certificate transparency by using CAs as authorities, although it may
be most beneficial in settings that present similar challenges to the ones
discussed in the introduction (i.e., in which objects are large and persistent
MitM attacks are a realistic threat).
\\~\\
\paragraphn{Archival node scalability.} The current design of \sysname requires
archival nodes to store all data, which as we have discussed in
Section~\ref{sec:debian} incurs a significant overhead.  There are likely many
alternative designs that alleviate these requirements, such as a \emph{sharded}
solution in which archival nodes store only the data for which they sufficient
space, and we leave an exploration of this space as an interesting open problem.

\section{Conclusion}\label{sec:conclusions}

We have proposed \sysname, a system that provides proactive transparency,
logarithmic scaling for auditors in the number of packages they have installed, 
and does not require the initial coordination of forming a Sybil-free set of 
nodes.    
We have demonstrated that, even for attackers that are capable of performing 
persistent man-in-the-middle attacks, compromising the integrity of the 
system requires millions of dollars in energy and hardware costs.  We also 
saw that \sysname could be applied today to the Debian software repository 
with relatively low overhead to existing infrastructure, and with no
changes or coordination required for any participant (even the Debian server) 
who does not wish to opt in.  

\section*{Acknowledgements}

Mustafa Al-Bassam is supported by a scholarship from the Alan Turing Institute,
and Sarah Meiklejohn is supported by EPSRC grant EP/N028104/1.

{\footnotesize
\bibliographystyle{abbrv}
\bibliography{contour-paper,misc}

\begin{thebibliography}{10}

\bibitem{mozillabt2017}
Security/binary transparency - mozillawiki, 2017.
\newblock \url{https://wiki.mozilla.org/Security/Binary_Transparency}.

\bibitem{hijacking-bitcoin}
M.~Apostolaki, A.~Zohar, and L.~Vanbever.
\newblock {Hijacking Bitcoin: Large-scale Network Attacks on Cryptocurrencies},
  2016.
\newblock \url{arxiv.org/abs/1605.07524}.

\bibitem{massimobartoletti2017}
M.~Bartoletti and L.~Pompianu.
\newblock {An analysis of Bitcoin OP\_RETURN metadata}.
\newblock In {\em 4th Workshop on Bitcoin and Blockchain Research}, 2017.

\bibitem{arpki}
D.~Basin, C.~Cremers, T.~H.-J. Kim, A.~Perrig, R.~Sasse, and P.~Szalachowski.
\newblock {ARPKI: Attack Resilient Public-Key Infrastructure}.
\newblock In {\em ACM CCS 2014}, pages 382--393, 2014.

\bibitem{ethiks}
J.~Bonneau.
\newblock {EthIKS: Using Ethereum to audit a CONIKS key transparency log}.
\newblock In {\em 3rd Workshop on Bitcoin and Blockchain Research}, 2016.

\bibitem{melissachase2016}
M.~Chase and S.~Meiklejohn.
\newblock {Transparency Overlays and Applications}.
\newblock In {\em ACM SIGSAC Conference on Computer and Communications
  Security}, 2016.

\bibitem{laurentchuat2015}
L.~Chuat, P.~Szalachowski, A.~Perrig, B.~Laurie, and E.~Messeri.
\newblock {Efficient Gossip Protocols for Verifying the Consistency of
  Certificate Logs}.
\newblock In {\em IEEE Conference on Communications and Network Security},
  2015.

\bibitem{xavierdecarnedecarnavalet2014}
X.~de~Carné~de Carnavalet and M.~Mannan.
\newblock Challenges and implications of verifiable builds for
  security-critical open-source software.
\newblock In {\em 30th Annual Computer Security Applications Conference}, 2014.

\bibitem{ct-eprint}
B.~Dowling, F.~G{\"u}nther, U.~Herath, and D.~Stebila.
\newblock {Secure Logging Schemes and {Certificate} {Transparency}}.
\newblock In {\em ESORICS 2016}, 2016.

\bibitem{continusec}
A.~Eijdenberg, B.~Laurie, and A.~Cutter.
\newblock {Verifiable Data Structures}, 2015.
\newblock
  \url{github.com/google/trillian/blob/master/docs/VerifiableDataStructures.pdf}.

\bibitem{sabaeskandarian2017}
S.~Eskandarian, E.~Messeri, J.~Bonneau, and D.~Boneh.
\newblock Certificate transparency with privacy.
\newblock {\em CoRR}, abs/1703.02209, 2017.

\bibitem{ittayeyal2014}
I.~Eyal and E.~G. Sirer.
\newblock Majority is not enough: Bitcoin mining is vulnerable.
\newblock In {\em Financial Cryptography and Data Security}, 2014.

\bibitem{cyrusfarivar2016}
C.~Farivar.
\newblock {Judge: Apple must help FBI unlock San Bernardino shooter's iPhone},
  2016.
\newblock
  \url{arstechnica.com/tech-policy/2016/02/judge-apple-must-help-fbi-unlock-san-bernardino-shooters-iphone/}.

\bibitem{certcoin}
C.~Fromknecht, D.~Velicanu, and S.~Yakoubov.
\newblock A decentralized public key infrastructure with identity retention.
\newblock IACR Cryptology ePrint Archive, Report 2014/803, 2014.
\newblock \url{eprint.iacr.org/2014/803.pdf}.

\bibitem{simongarfinkel1996}
S.~Garfinkel.
\newblock {\em PGP: Pretty Good Privacy}.
\newblock O'Reilly Media, Sebastopol, CA, USA, 1st edition, 1996.

\bibitem{gervais-eclipse}
A.~Gervais, H.~Ritzdorf, G.~Karame, and S.~Capkun.
\newblock {Tampering with the Delivery of Blocks and Transactions in Bitcoin}.
\newblock In {\em ACM CCS 2015}, 2015.

\bibitem{dangoodin2012}
D.~Goodin.
\newblock {``Flame'' malware was signed by rogue Microsoft certificate}, 2012.
\newblock
  \url{arstechnica.com/security/2012/06/flame-malware-was-signed-by-rogue-microsoft-certificate/}.

\bibitem{heilman-eclipse}
E.~Heilman, A.~Kendler, A.~Zohar, and S.~Goldberg.
\newblock {Eclipse Attacks on Bitcoin's Peer-to-Peer Network}.
\newblock In {\em USENIX Security 2015}, 2015.

\bibitem{aki}
T.~H.-J. Kim, L.-S. Huang, A.~Perrig, C.~Jackson, and V.~Gligor.
\newblock {Accountable key infrastructure {(AKI)}: a proposal for a public-key
  validation infrastructure}.
\newblock In {\em WWW 2013}, pages 679--690, 2013.

\bibitem{bitcoin-cosi}
E.~K. Kogias, P.~Jovanovic, N.~Gailly, I.~Khoffi, L.~Gasser, and B.~Ford.
\newblock {Enhancing Bitcoin Security and Performance with Strong Consistency
  via Collective Signing}.
\newblock In {\em USENIX Security 2016}, 2016.

\bibitem{ietf-ct}
B.~Laurie, A.~Langley, and E.~Kasper.
\newblock Certificate transparency, 2013.

\bibitem{johnleyden2011}
J.~Leyden.
\newblock {Inside 'Operation Black Tulip': DigiNotar hack analysed}, 2011.
\newblock \url{www.theregister.co.uk/2011/09/06/diginotar_audit_damning_fail/}.

\bibitem{ikp}
S.~Matsumoto and R.~M. Reischuk.
\newblock {IKP: Turning a PKI Around with Blockchains}.
\newblock IACR Cryptology ePrint Archive, Report 2016/1018, 2016.
\newblock \url{eprint.iacr.org/2016/1018}.

\bibitem{coniks}
M.~S. Melara, A.~Blankstein, J.~Bonneau, E.~W. Felten, and M.~J. Freedman.
\newblock {{CONIKS}: Bringing Key Transparency to End Users}.
\newblock In {\em USENIX Security 2015}, 2015.

\bibitem{satoshinakamoto2008}
S.~Nakamoto.
\newblock {Bitcoin: A Peer-to-Peer Electronic Cash System}, 2008.
\newblock \url{bitcoin.org/bitcoin.pdf}.

\bibitem{nikitin2017}
K.~Nikitin, E.~Kokoris-Kogias, P.~Jovanovic, N.~Gailly, L.~Gasser, I.~Khoffi,
  J.~Cappos, and B.~Ford.
\newblock {CHAINIAC}: Proactive software-update transparency via collectively
  signed skipchains and verified builds.
\newblock In {\em 26th {USENIX} Security Symposium ({USENIX} Security 17)},
  pages 1271--1287, Vancouver, BC, 2017. {USENIX} Association.

\bibitem{linusnordberg2015}
L.~Nordberg, D.~Gillmor, and T.~Ritter.
\newblock {Gossiping in CT}, 2016.
\newblock \url{tools.ietf.org/html/draft-ietf-trans-gossip-03}.

\bibitem{enhanced-ct}
M.~D. Ryan.
\newblock {Enhanced Certificate Transparency and End-to-end Encrypted Mail}.
\newblock In {\em NDSS 2014}, 2014.

\bibitem{atulsingh2006}
A.~Singh, T.-W.~J. Ngan, P.~Druschel, and D.~S. Wallach.
\newblock Eclipse attacks on overlay networks: Threats and defenses.
\newblock In {\em IEEE Conference on Computer Communications}, 2006.

\bibitem{ewasyta2015}
E.~Syta, I.~Tamas, D.~Visher, D.~I. Wolinsky, P.~Jovanovic, L.~Gasser,
  N.~Gailly, I.~Khoffi, and B.~Ford.
\newblock {Keeping Authorities ``Honest or Bust'' with Decentralized Witness
  Cosigning}.
\newblock In {\em IEEE Symposium on Security and Privacy (``Oakland'')}, 2016.

\bibitem{alintomescu2016}
A.~Tomescu and S.~Devadas.
\newblock {Catena: Efficient Non-equivocation via Bitcoin}.
\newblock In {\em IEEE Symposium on Security and Privacy (``Oakland'')}, 2017.

\end{thebibliography}
}

\appendix

\section{Cost of a Split-View Attack}\label{sec:appendix}

To support our argument in Section~\ref{sec:security-proof} about the
infeasibility of carrying out a split-view attack, we provide here more concrete
estimates for the associated costs of the attack.  These are rough estimates, as
they make assumptions about certain properties (e.g., electricity costs and
choice of mining hardware) that are not guaranteed to hold in practice.
We are not aware of any previous literature considering the costs
of eclipse attacks on Bitcoin nodes, so we consider these estimates (even if
rough) to be important.

We first calculate the cost to mine a single block, and then analyze the cost of
performing a split-view attack in the case where the adversary is able to
perform an eclipse attack and where it cannot.

\paragraph{Cost to mine a single block.}
The probability of a miner finding a valid block after each hashing attempt is
$\frac{2^{16}-1}{2^{48}D}$, where $D$ is the periodically adjusted difficulty
of the network.  For a miner to mine a block then, they must make on
average $\frac{2^{48}D}{2^{16}-1}$ hashing attempts.  The total electricity
cost ($C$) of mining a block is thus
\begin{equation}\label{eqn:mining}
C = \frac{2^{48}D}{2^{16}-1}\cdot J\cdot E,
\end{equation}
where $J$ is the number of joules required per hashing attempt, and $E$ is
the electricity cost of one joule. As of December 2017, the most
energy-efficient Bitcoin mining hardware is the Antminer S9, which has an
energy cost of $9.82\cdot10^{-11}$ joules per
hash,\footnote{\url{en.bitcoin.it/wiki/Mining_hardware_comparison}}
and the average retail price of one kilowatt hour in the US is
0.10 USD.\footnote{\url{www.eia.gov/electricity/state/}}  The
cost per joule, $E$, is therefore
$\frac{0.10}{1000\cdot 60\cdot 60} = 2.8\cdot 10^{-8}$ USD.  As of
December 2017, the Bitcoin mining difficulty ($D$) is 1,347,001,430,558.
Plugging these numbers into Equation~\ref{eqn:mining}, the total electricity
cost to mine a block, using the most efficient hardware and assuming standard
electricity costs, is thus 15,908 USD.

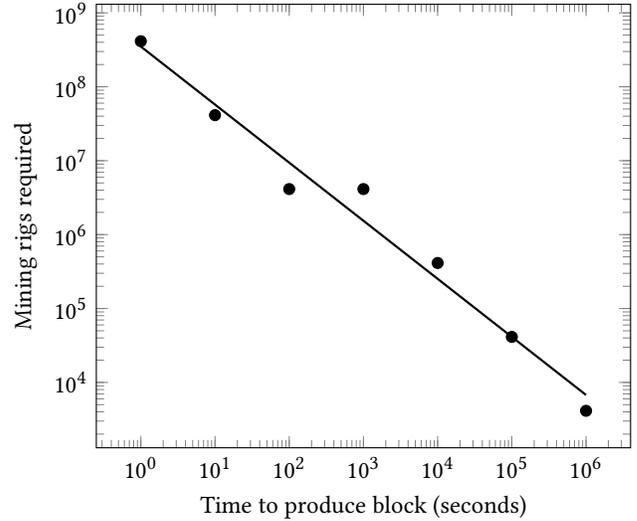
\begin{figure}
\resizebox {\columnwidth} {!} {
\begin{tikzpicture}
\begin{axis}[
xmin=0,
xmode=log,
ymode=log,
xlabel={Time to produce block (seconds)},
ylabel={Mining rigs required},
]
\addplot [only marks, mark = *] table {rig-number-table.dat};
\addplot [thick] table[
y={create col/linear regression={y=Y}}
] 
{rig-number-table.dat};
\end{axis}
\end{tikzpicture}
}
\caption{The number of Antminer S9 rigs required to produce
blocks under a certain time limit.}
\label{fig:mining}
\end{figure}

To also take hardware costs into account, the number of mining
rigs $N$ needed to mine a block in $S$ seconds is
\begin{equation}\label{eqn:number-rigs}
N=\frac{(\frac{2^{48}D}{2^{16}-1})}{H\cdot S},
\end{equation}
where $H$ is the number of hashes that the mining rig is capable of
calculating per second.  This formula is graphed in Figure~\ref{fig:mining}
for the Antminer S9 rig, which is capable of calculating 14
terahashes per second and has a retail cost of 2,400
USD.\footnote{\url{www.amazon.com/Antminer-S9-0-10W-Bitcoin-Miner/dp/B01GFEOV0O}}
We use these formulas to estimate the cost of split-view attacks in the
following analysis.
\\~\\
\paragraphn{Using eclipse attacks.} If an eclipse attack is possible, an
adversary can launch a successful split-view attack solely by mining $k$ blocks
at its own pace, where $k$ is the number of blocks the auditor requires to be
mined after a block containing a given commitment in order to consider that
commitment as valid. (It is standard in most Bitcoin wallets to use $k=6$.)

Using our rough estimates above, it would cost the adversary 15,908 USD in
electricity costs to mine a block, or 95,448 USD for $k=6$.  The hardware costs
depends on how much time the adversary needs to conduct the attack, or how long
they are able to continue their man-in-the-middle attack on the auditor. If\dash
as a conservative number\dash the adversary wants to conduct the attack within a
week, it must mine a block every 1.4 days to produce 6 blocks, which requires
3,417 mining rigs at a hardware cost of 8,200,800 USD.  This brings the total
cost of the attack to 8.3M USD.
Moreover, this attack is also fundamentally targeted: if the
adversary wants to later compromise previously non-eclipsed auditors, it must
mine a new set of blocks (assuming these auditors have more up-to-date blocks)
and pay the electricity costs again.  Even for an adversary with few financial
constraints, this makes it significantly more difficult to conduct such an
attack on a wide scale.

Furthermore, if the adversary takes 1.4 days to mine a block, or in general
the auditor sees no new blocks until long after the expected 10-minute
interval, it may assume that an eclipse attack is being performed.  We can
thus greatly increase the cost of the attack by adding simple
checks to the auditor to ensure that there is a maximum interval between
blocks.
If we generously set such a check to require a maximum of 3 hours between
blocks, then a total of 38,263 mining rigs are required at a cost of
91.8M USD.

In addition, the blocks must still follow the same difficulty level as honest
blocks, so by mining these only in the eclipsed view of the network the
adversary is not only expending the energy needed to do so but is also
forfeiting the mining reward associated with them. As of December 5 2017, the
Bitcoin mining reward is 12.5 bitcoins, or roughly 145,250 USD, so for $k=6$
the adversary must additionally forfeit 871,500 USD.
\\~\\
\paragraphn{Ignoring eclipse attacks.} To perform a split-view attack without an
eclipse attack, an adversary must fork the Bitcoin blockchain, which na{\"i}vely
requires control of 51\% of the network's mining power.

As of December 5 2017, the total hashing power of the Bitcoin network was
11,918,845 terahashes per
second.\footnote{\url{blockchain.info/charts/hash-rate}}
Conducting a 51\% attack would therefore require the adversary to be able to
compute more than 11,918,845 terahashes per second.  Per
hour, the total electricity cost would be $11918845\cdot10^{12}\cdot
3600\cdot J\cdot E$, or\dash using our earlier estimates for $J$ and $E$\dash
117,979 USD per hour.  In terms of hardware costs, if we use the figures for
the Antminer S9 from before, the total number of mining rigs required would be
greater than $\frac{11918845\cdot10^{12}}{14\cdot10^{12}} = 851346$, at a
total cost of 2043M USD.

While more sophisticated attacks, such as selfish mining~\cite{ittayeyal2014},
have proposed strategies that fork the blockchain using only 25\% of the mining
power, this would still require an investment of hundreds of millions of
dollars.  Such an attack would furthermore be highly visible, as the blockchain
is regularly monitored for forks.

\end{document}